\documentclass{svproc}
\usepackage{url}

\usepackage[
backend=biber,
style=numeric,
autocite=footnote,
sorting=nty,
sortcase=false,
url=true,
date=year,
eprint=false,
hyperref=auto,
maxbibnames=99,
giveninits=true,
isbn=false
]{biblatex}
\usepackage{amsmath}
\usepackage{amsfonts}
\usepackage{graphicx}

\begin{document}
%
\title{Dynamic mode decomposition of noisy flow data}
%
%
\author{Andre Weiner\inst{1} \and Janis Geise\inst{1}}
%
%
%
%
\institute{$^1$Technical University of Dresden, Institute of Fluid Mechanics,\\
\email{andre.weiner@tu-dresden.de, janis.geise@tu-dresden.de},\\
}


\maketitle              

\begin{abstract}
	
Dynamic mode decomposition (DMD) is a popular approach to analyzing and modeling fluid flows.
In practice, datasets are almost always corrupted to some degree by noise.
The vanilla DMD is highly noise-sensitive, which is why many algorithmic extensions for improved robustness exist.
We introduce a flexible optimization approach that merges available ideas for improved accuracy and robustness.
The approach simultaneously identifies coherent dynamics and noise in the data.
In tests on the laminar flow past a cylinder, the method displays strong noise robustness and high levels of accuracy.
 
\end{abstract}
\section{Introduction}
\label{sec:intro}

Modal decomposition techniques have become a standard for the snapshot-based analysis and modeling of complex fluid flows.
Taira et al. \cite{taira2017} provide a comparative overview of available approaches.
The \textit{dynamic mode decomposition} (DMD)by P. Schmid \cite{schmid2010,tu2014} is a popular decomposition technique that combines elements of dimensionality reduction via the \textit{proper orthogonal decomposition} (POD) and the \textit{discrete Fourier transformation} (DFT).
Similar to the DFT, the DMD yields frequency-amplitude pairs that are relatively straightforward to interpret.
In contrast to the DFT, the DMD spectrum is not fixed to predefined frequency bins, and the POD dimensionality reduction allows a selective truncation of the spectrum.
The truncation further improves interpretability and mitigates the influence of noise.

DMD dynamics may grow or shrink over time.
This additional degree of freedom can sometimes introduce undesirable effects in the presence of noise \cite{dawson2016}.
A number of DMD extensions exist to improve noise robustness \cite{schmid2022}.
While each extension typically has some positive effect on accuracy and robustness, it is not straightforward to combine their favorable aspects due to algorithmic differences in the optimization.

Our contribution introduces an optimization approach based on automatic differentiation and gradient descent that enables us to merge the strengths and core ideas of various DMD variants.
In the remainder of this article, we revisit theoretical aspects of selected DMD variants in section \ref{sec:theory}, introduce an enhanced optimization approach to identify coherent dynamics and noise in section \ref{sec:new_dmd}, describe our numerical experiments in section \ref{sec:method}, and present comparative results for the flow past a cylinder in section \ref{sec:results}.
The code repository accompanying this article \cite{weiner2024} allows reproducing the benchmarks.
The improved DMD variant is available via an open-source library \cite{weiner2021}.

\section{Theoretical background}
\label{sec:theory}
\subsection{Dimensionality reduction}

A sample of the flow state at time $t_n = n\Delta t$ may be expressed as vector $\mathbf{x}_n$, $\mathbf{x}_n \in \mathbb{R}^M$.
We assume that $N$ consecutive state vectors sampled at constant intervals $\Delta t$ are arranged in a data matrix $\mathbf{M} = \left[\mathbf{x}_1, \mathbf{x}_2, \ldots, \mathbf{x}_N \right]^T$, $\mathbf{M} \in \mathbb{R}^{M\times N}$.
In the experimental or numerical analysis of fluid dynamics, the raw state vector is often high-dimensional, i.e., $M$ can take on values in the millions or billions.
However, the temporal evolution of most entries in $\mathbf{x}$ is usually highly correlated such that the main flow dynamics may be accurately described in terms of a reduced state vector $\tilde{\mathbf{x}}_n$, $\tilde{\mathbf{x}}_n \in \mathbb{R}^r$ and $r\ll M$.
For the DMD, the dimensionality reduction is achieved by projecting the data matrix onto the first $r$ POD modes. The POD modes may be computed employing a \textit{singular value decomposition} (SVD):
\begin{equation}
    \mathbf{M} \approx \mathbf{U}_r\mathbf{\Sigma}_r\mathbf{V}_r^T,\quad \tilde{\mathbf{M}} = \mathbf{U}_r^T\mathbf{M} = \mathbf{\Sigma}_r\mathbf{V}_r^T,
\end{equation}
where $\mathbf{U}_r\in \mathbb{R}^{M\times r}$, $\mathbf{V}_r \in \mathbb{R}^{N \times r}$, $\mathbf{\Sigma}_r \in \mathbb{R}^{r\times r}$, and  $\tilde{\mathbf{M}} \in \mathbb{R}^{r\times N}$.
The number of POD basis vectors should be sufficiently large to capture the main flow dynamics.
Several approaches exist to choose suitable values for $r$, e.g., based on a threshold for the explained variance or the decay of the singular values; see references \cite{askham2018,weiner2023a} for a more detailed discussion.
In the following, we discuss selected DMD variants and related approaches in terms of the reduced state vector $\tilde{\mathbf{x}}_n$ and the corresponding data matrix $\tilde{\mathbf{M}}$.

\subsection{Dynamic mode decomposition}

The DMD assumes a linear relationship between two consecutive states $ \tilde{\mathbf{x}}_n$ and $\tilde{\mathbf{x}}_{n+1}$ such that \cite{schmid2010}:
\begin{equation}
	\label{eq:dmd_map}
	\tilde{\mathbf{x}}_{n+1} = \tilde{\mathbf{A}} \tilde{\mathbf{x}}_{n}.
\end{equation}
To also handle dynamics that are not perfectly linear, the operator $\tilde{\mathbf{A}} \in \mathbb{R}^{r\times r}$ is defined in a least-square sense between pairs of snapshots arranged in two data matrices $\tilde{\mathbf{X}}$ and $\tilde{\mathbf{Y}}$:
\begin{equation}
	\label{eq:least_squares}
	\underset{\tilde{\mathbf{A}}}{\mathrm{argmin}}\left|\left| \tilde{\mathbf{Y}}-\tilde{\mathbf{A}}\tilde{\mathbf{X}} \right|\right|_2
	=\tilde{\mathbf{Y}}\tilde{\mathbf{X}}^\dagger.
\end{equation}
When using all snapshots in $\tilde{\mathbf{M}}$, the matrices $\tilde{\mathbf{X}}$ and  $\tilde{\mathbf{Y}}$ consist of the first and the last $N-1$ projected snapshots, respectively.
The upper script $\dagger$ denotes the pseudoinverse (Moore-Penrose inverse).

The eigendecomposition of the linear operator, $\tilde{\mathbf{A}}=\tilde{\mathbf{\Phi}}\tilde{\mathbf{\Lambda}}\tilde{\mathbf{\Phi}}^{-1}$ with $\tilde{\mathbf{\Phi}}, \tilde{\mathbf{\Lambda}} \in \mathbb{C}^{r\times r}$, makes the dynamics interpretable and also yields an efficient reduced-order model that enables predictions at any future time $t_n$ given an initial condition $\tilde{\mathbf{x}}_1$ \cite{askham2018}:
\begin{equation}
	\label{eq:rec_n}
	\tilde{\mathbf{x}}_n \approx \tilde{\mathbf{\Phi}}\tilde{\mathbf{\Lambda}}^{n-1}\tilde{\mathbf{b}},
\end{equation}
with $\tilde{\mathbf{b}} = \tilde{\mathbf{\Phi}}^{-1}\tilde{\mathbf{x}}_1$. Employing relation \eqref{eq:rec_n} for all $N$ time instances allows recovering the projected data matrix:
\begin{equation}
	\label{eq:rec_full}
	\tilde{\mathbf{M}} \approx \tilde{\mathbf{\Phi}}\tilde{\mathbf{D}}_\mathbf{b}\tilde{\mathbf{V}}_\mathbf{\lambda},
\end{equation}
where $\tilde{\mathbf{D}}_\mathbf{b}=\mathrm{diag}(\tilde{\mathbf{b}})$ and $\tilde{\mathbf{V}}_{\mathbf{\lambda}} = \left[\tilde{\mathbf{\lambda}}^0, \tilde{\mathbf{\lambda}}^1, \ldots, \tilde{\mathbf{\lambda}}^{N-1}\right]^T$.
Based on the approximately reconstructed data matrix \eqref{eq:rec_full}, the normalized reconstruction error in the reduced space may be defined as:
\begin{equation}
	\label{eq:e_rec}
	E_\mathrm{rec} = \frac{||\tilde{\mathbf{M}} - \tilde{\mathbf{\Phi}}\tilde{\mathbf{D}}_\mathbf{b}\tilde{\mathbf{V}}_\mathbf{\lambda}||_F}{||\tilde{\mathbf{M}}||_F}.
\end{equation}
For completeness it remains to be noted that the so-called \textit{projected} DMD modes \cite{tu2014} and the full data matrix may be reconstructed as $\mathbf{\Phi} \approx \mathbf{U}_r \tilde{\mathbf{\Phi}}$ and $\mathbf{M} \approx \mathbf{U}_r \tilde{\mathbf{M}}$, respectively.

\subsection{Forward-backward consistency}

It is well-known that noise-corrupted snapshots bias the identified operator towards decaying dynamics \cite{dawson2016}.
When predicting backward in time, i.e., $\tilde{\mathbf{x}}_{n} = \tilde{\mathbf{A}}^{-1} \tilde{\mathbf{x}}_{n+1}$, the trend is inverted.
While spuriously decaying dynamics impact the prediction accuracy, they do not jeopardize the numerical stability since the influence of affected modes vanishes over time.
The same statement does not hold for spuriously growing dynamics, which can lead to catastrophic prediction errors and numerical overflow.
This potential inconsistency between forward and backward predictions provides an additional constraint for the operator definition \cite{azencot2019}:
\begin{equation}
    \label{eq:fb_cons}
    \underset{\tilde{\mathbf{A}}}{\mathrm{argmin}}\left|\left| \tilde{\mathbf{Y}}-\tilde{\mathbf{A}}\tilde{\mathbf{X}} \right|\right|_F +
    \left|\left| \tilde{\mathbf{X}}-\tilde{\mathbf{A}}^{-1}\tilde{\mathbf{Y}} \right|\right|_F.
\end{equation}
Due to the operator's inverse, the optimization problem \eqref{eq:fb_cons} becomes nonlinear and non-convex.
However, introducing the auxiliary operator $\tilde{\mathbf{B}}=\tilde{\mathbf{A}}^{-1}$ makes the optimization amenable to the \textit{alternating direction method of multipliers} \cite{azencot2019}:
\begin{equation}
    \label{eq:cdmd}
    \underset{\tilde{\mathbf{A}}, \tilde{\mathbf{B}}}{\mathrm{argmin}}\left|\left| \tilde{\mathbf{Y}}-\tilde{\mathbf{A}}\tilde{\mathbf{X}} \right|\right|_F +
    \left|\left| \tilde{\mathbf{X}}-\tilde{\mathbf{B}}\tilde{\mathbf{Y}} \right|\right|_F\quad
    \text{s.t.}\quad \tilde{\mathbf{A}}\tilde{\mathbf{B}}=\tilde{\mathbf{I}},\ \tilde{\mathbf{B}}\tilde{\mathbf{A}}=\tilde{\mathbf{I}}.
\end{equation}
This approach to ensuring forward-backward consistency results in the \textit{consistent} DMD (CDMD) and was first presented by Azencot et al. \cite{azencot2019}.


\subsection{Error amplification and optimal mode amplitudes}

Since the data-fitted linear operator is only an approximation, it is insightful to analyze how errors in the corresponding eigendecomposition affect the prediction of future states.
According to relation \eqref{eq:rec_n}, the prediction error is proportional to errors in the eigenvectors $\tilde{\mathbf{\Phi}}$ and mode amplitudes $\tilde{\mathbf{b}}$, while discrepancies in the eigenvalues $\tilde{\mathbf{\lambda}}$ are amplified exponentially over time.
Including the error amplification in the operator definition provides another powerful constraint.
In the \textit{optimized} DMD (ODMD) by Askham and Kutz \cite{askham2018}, this idea is incorporated using relation \eqref{eq:rec_full} as follows:
\begin{equation}
	\label{eq:optDMD}
	\underset{\tilde{\mathbf{\lambda}},\tilde{\mathbf{\Phi}}_\mathbf{b}}{\mathrm{argmin}}\left|\left| \tilde{\mathbf{M}}-\tilde{\mathbf{\Phi}}_\mathbf{b}\tilde{\mathbf{V}}_{\mathbf{\lambda}} \right|\right|_F,
\end{equation}
where $\tilde{\mathbf{\Phi}}_\mathbf{b} = \tilde{\mathbf{\Phi}}\tilde{\mathbf{D}}_\mathbf{b}$ are rescaled eigenvectors.
Askham and Kutz employ \textit{variable projection} \cite{askham2018} to obtain the optimized eigenvectors and eigenvalues.
The mode amplitudes are recovered by scaling the eigenvectors $\tilde{\mathbf{\phi}}_{b,i}$ to unit length, i.e., $\tilde{b}_i = \left|\tilde{\mathbf{\phi}}_{b,i}\right|$ and $\tilde{\mathbf{\phi}}_i = \tilde{\mathbf{\phi}}_{b,i} / \tilde{b}_i$. 

While posing definition \eqref{eq:optDMD} in terms of the scaled eigenvectors is necessary to employ variable projection, there are more subtle consequences of this choice.
Since $\tilde{\mathbf{b}}$ is a result of the optimization, the initial condition is learned implicitly as well since $\tilde{\mathbf{\Phi}}\tilde{\mathbf{b}} = \tilde{\mathbf{x}}_1$.
In the presence of noise, the implicitly learned initial condition significantly improves the reconstruction accuracy and robustness of the ODMD.
The learned initial condition might be interpreted as an enhanced or denoised version of the original snapshot.

\subsection{Combining multiple constraints and noise identification}
\label{sec:new_dmd}

In the following, we introduce a definition of the linear operator that combines and extends the powerful constraints and concepts introduced in the previous sections.
First, we assume that the observed snapshot $\tilde{\mathbf{x}}_n$ is a noisy version of the true state $\hat{\mathbf{x}}_n$:
\begin{equation}
    \label{eq:x_noise}
    \tilde{\mathbf{x}}_n = \hat{\mathbf{x}}_n + \tilde{\mathbf{n}}_n,\quad
    \tilde{\mathbf{M}} = \hat{\mathbf{M}} + \tilde{\mathbf{N}},
\end{equation}
where $\tilde{\mathbf{N}}\in \mathbb{R}^{r\times N}$ is a matrix holding the noise vectors $\tilde{\mathbf{n}}_n$.
This idea has been employed previously in the context of \textit{sparse identification of nonlinear dynamics} (SINDy) \cite{kaheman2022}.
In analogy to expression \eqref{eq:rec_n}, predictions of future or past noise-free states at time $t_{n+h} = (n+h)\Delta t$ may be computed as:
\begin{equation}
    \label{eq:rec_fb}
    \hat{\bar{\mathbf{x}}}_{n+h} = \tilde{\mathbf{\Phi}}\tilde{\mathbf{\Lambda}}^{n+h-1}\tilde{\mathbf{\Phi}}^{-1} \hat{\mathbf{x}}_n,
\end{equation}
where $\hat{\mathbf{x}}_n = \tilde{\mathbf{x}}_n-\tilde{\mathbf{n}}_n$.
Note that equation \eqref{eq:rec_fb} works for positive and negative values of $h$.
The eigenvalue matrix raised to a negative power denotes the inverse of the matrix exponential.
Also note that we use the notation $\hat{\mathbf{x}}_{n}$ and $\hat{\bar{\mathbf{x}}}_{n}$ to distinguish between noise-free states obtained from equation \eqref{eq:x_noise} and \eqref{eq:rec_fb}, respectively.
Based on equation \eqref{eq:rec_fb}, we define the forward and backward loss, $L_f$ and $L_b$, as:
\begin{equation}
    \label{eq:loss_ocdmd}
    L_f = \sum\limits_{n=1}^{N-H}\sum\limits_{h=1}^{H} \left|\tilde{\mathbf{x}}_{n+h} - \hat{\bar{\mathbf{x}}}_{n+h}\right|,\quad
    L_b = \sum\limits_{n=H+1}^{N}\sum\limits_{h=1}^{H} \left|\tilde{\mathbf{x}}_{n-h} - \hat{\bar{\mathbf{x}}}_{n-h}\right|,
\end{equation}
where the prediction horizon $H$ is a user-defined parameter in the range $1\leq H \leq N-1$.
The final minimization problem in terms of eigenvectors, eigenvalues, and noise reads:
\begin{equation}
	\label{eq:OCDMD}
	\underset{\tilde{\mathbf{\lambda}},\tilde{\mathbf{\Phi}}, \tilde{\mathbf{N}}}{\mathrm{argmin}} \left(L_f + L_b\right).
\end{equation}
We chose to formulate the loss in terms of the difference $\tilde{\mathbf{x}}_{n} - \hat{\bar{\mathbf{x}}}_{n}$ rather than $\hat{\mathbf{x}}_{n} - \hat{\bar{\mathbf{x}}}_{n}$, which is in contrast to the SINDy approach \cite{kaheman2022}.
Employing the difference $\hat{\mathbf{x}}_{n} - \hat{\bar{\mathbf{x}}}_{n}$ can lead to unwanted solutions for high noise levels, i.e., $\tilde{\mathbf{x}}_{n}=\tilde{\mathbf{n}}_{n}$, if the magnitude of the noise vectors is not penalized; refer to reference \cite{kaheman2022} for a detailed discussion.
Our numerical experiments showed that the difference $\tilde{\mathbf{x}}_{n} - \hat{\bar{\mathbf{x}}}_{n}$ regularizes the optimization while leading to a comparable reconstruction accuracy.
The noise correction enters the loss function only via the initial condition.
However, if $H<N/2$, the loss contains all noise vectors since all snapshots are used at least once as the initial condition.
For $H=N-1$, the forward loss is equivalent to the ODMD formulation \eqref{eq:optDMD}.
We want to stress that learning the noise is essential to recovering the ODMD's high reconstruction accuracy for low noise levels.
For $H=1$, the loss function is similar to the CDMD with additional noise identification.
An additional discussion of the prediction horizon follows in section \ref{sec:results}.

The minimization problem \eqref{eq:OCDMD} is highly nonlinear and non-convex.
However, thanks to the powerful toolchains developed by the deep learning community,
the minimization can be performed with reasonable effort.
We employ automatic differentiation, the ADAM gradient descent algorithm, and a plateau-based learning rate scheduler.
A detailed description of all implementation details would exceed the scope of the article.
Please refer to the publicly available source code for more details \cite{weiner2024,weiner2021}.

The optimization approach has several benefits we would like to mention briefly.
First, the loss \eqref{eq:loss_ocdmd} may be evaluated on subsets of the available initial conditions.
Splitting the available data can be useful for efficient \textit{batch updates} as well as \textit{early stopping} \cite{weiner2023b}.
The latter technique avoids an over-adjustment of the free DMD parameters ($\tilde{\mathbf{\Phi}}$ and $\tilde{\mathbf{\Lambda}}$) to the noise in the data.
Finally, we note that multiple datasets may be merged in the optimization, which would not be possible with the original ODMD.

\section{Numerical experiments}
\label{sec:method}

\begin{figure}[htbp]
	\begin{center}
		\includegraphics[width=0.85\textwidth]{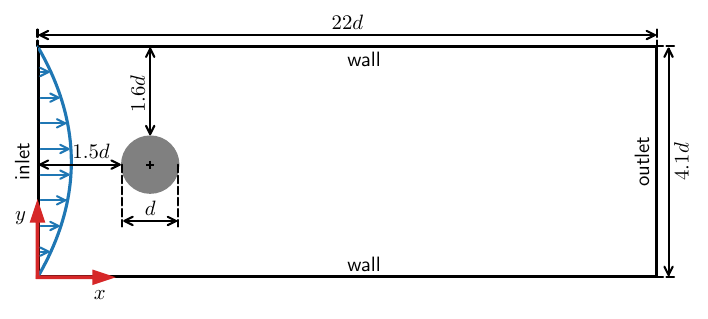}
		\caption{Simulation domain of the flow past a cylinder; based on \cite{schaefer1996}.}
		\label{fig:setup_cylinder}
	\end{center}
\end{figure}

We perform numerical tests on the incompressible 2D flow past a cylinder at $Re=dU_\mathrm{max}/\nu = 100$ ($d$ - cylinder diameter, $U_\mathrm{max}$ - inlet speed at $y=2d$, $\nu$ - kinematic viscosity).
Figure \ref{fig:setup_cylinder} shows a sketch of the domain.
A parabolic velocity profile is applied at the inlet, while the velocity gradient at the outlet is set to zero.
The velocity at all other boundaries is set to zero.
The pressure at the outlet is fixed, and at all other boundaries, the pressure gradient is set to zero.
The finite volume mesh consists of approximately $4\times 10^4$ control volumes.
The developed flow exhibits periodic vortex shedding at the cylinder.
We analyze $N=526$ snapshots of the plane-normal vorticity field comprising two complete cycles of vortex shedding.

\begin{figure}[htbp]
	\begin{center}
		\includegraphics[width=0.85\textwidth]{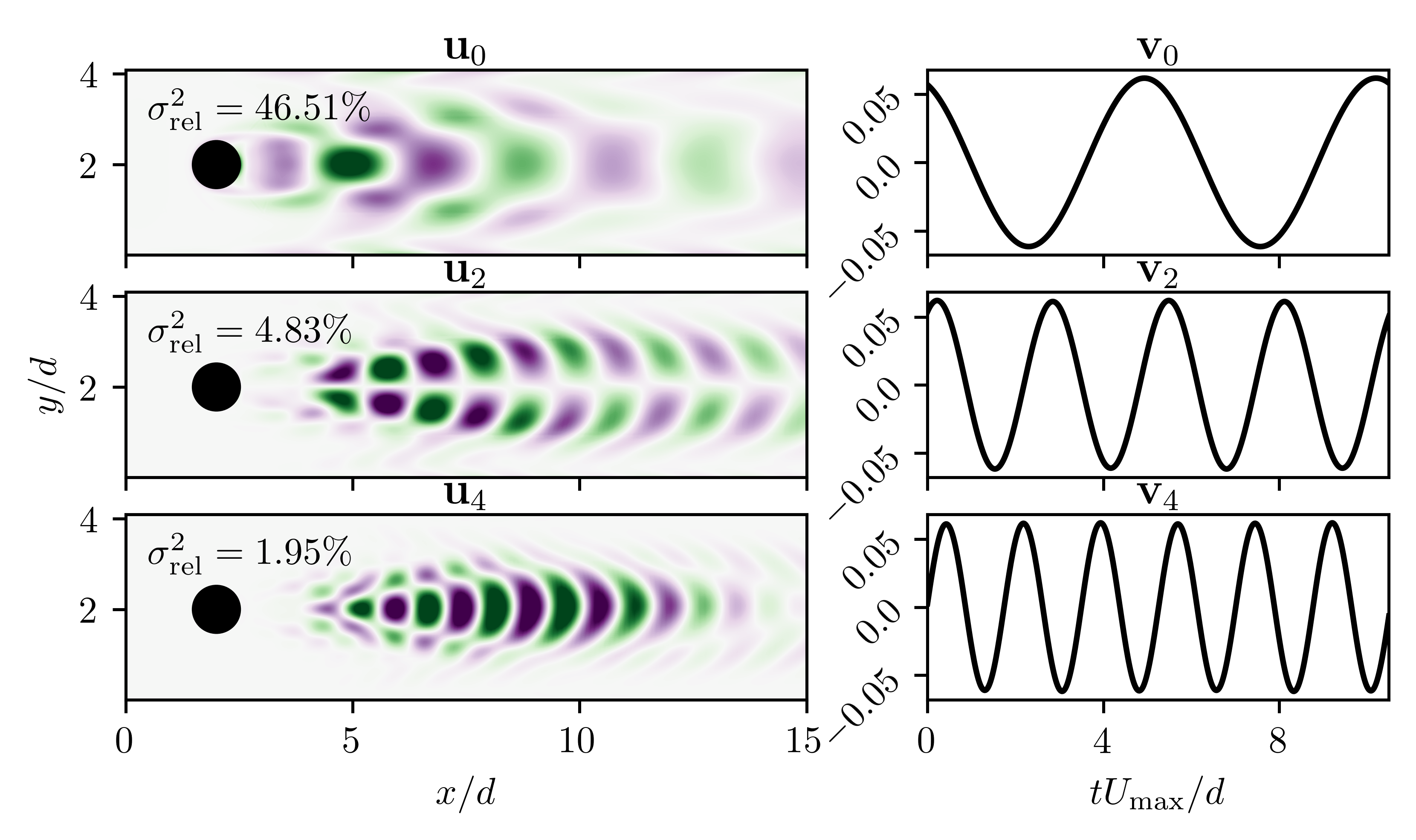}
		\caption{Dominant POD modes and normalized time coefficients; the contours are scaled to the range $\left[-1, 1\right]$ (magenta to green).}
		\label{fig:pod}
	\end{center}
\end{figure}

An SVD analysis of the mean-subtracted snapshots shows that the first six POD modes explain more than $99\%$ of the temporal variance \cite{weiner2024}.
Figure \ref{fig:pod} shows every second POD mode $\mathbf{u}_i$ and the corresponding right-singular vector $\mathbf{v}_i$.
The odd POD modes and coefficients are qualitatively similar and omitted for brevity.
The right-singular vectors are depicted as temporal signals since they correspond to the normalized POD coefficients.
Figure \ref{fig:pod} also shows the relative variance $\sigma^2_\mathrm{rel}$ explained by each mode-coefficient pair.

Since the leading vortex shedding mode together with the first two harmonics explain nearly all of the variance, we set $r=6$ such that $\tilde{\mathbf{M}}\in \mathbb{R}^{6\times 526}$.
To benchmark and compare different DMD flavors regarding their robustness to noise, we add noise sampled from a normal distribution to the mode coefficients. 
The normal distribution has zero mean and a standard deviation scaled in proportion to the temporal standard deviation of each mode coefficient such that the signal-to-noise ratio (SNR) is constant across all coefficients.
For the coefficient $\tilde{x}_i$ the noise level $l$ is defined as:
\begin{equation}
    \label{eq:snr}
    l = \mathrm{std}(\tilde{n}_i) / \mathrm{std}(\tilde{x}_i) \times 100\%.
\end{equation}
We quantify the noise identified by our DMD variant as:
\begin{equation}
    \label{eq:e_noise}
    E_\mathrm{noise} = \frac{||\tilde{\mathbf{N}}_\mathrm{true} - \tilde{\mathbf{N}}||_F}{||\tilde{\mathbf{N}}_\mathrm{true}||_F},
\end{equation}
where $\tilde{\mathbf{N}}_\mathrm{true}$ and $\tilde{\mathbf{N}}$ denote the prescribed and learned noise matrices, respectively.

\begin{figure}[htbp]
	\begin{center}
		\includegraphics[width=0.85\textwidth]{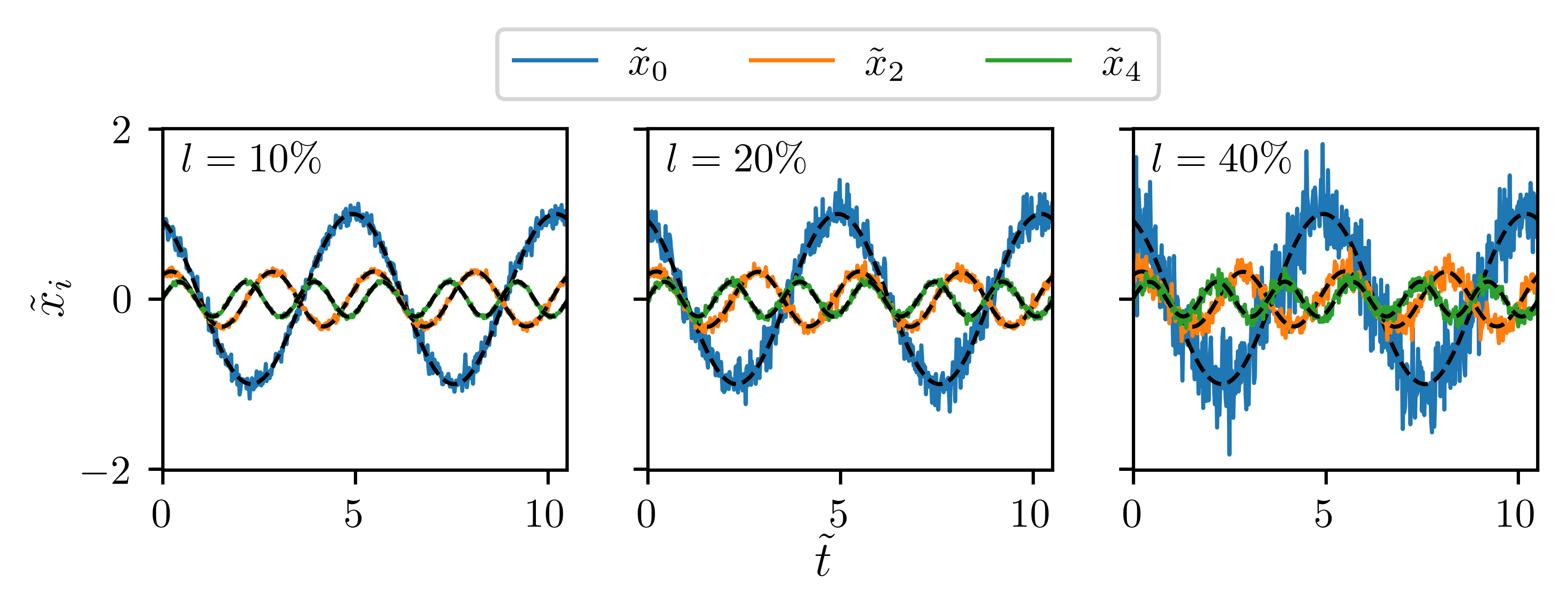}
		\caption{Time coefficients subjected to varying levels of noise; the dashed line indicates the ground truth; $\tilde{t} = tU_\mathrm{max}/d$.}
		\label{fig:noisy_data}
	\end{center}
\end{figure}

Figure \ref{fig:noisy_data} shows the even time coefficients subjected to the noise levels investigated in the next section.
Note that we could have also added the noise before the dimensionality reduction.
However, the SNR of the coefficients would be harder to quantify since the POD subspace projection removes some of the noise.

\section{Results}
\label{sec:results}

All numerical experiments are repeated for $20$ different noise samples.
We report results obtained with single-precision floating point numbers.
Switching to double precision did not show any noticeable difference.

\begin{figure}[htbp]
	\begin{center}
		\includegraphics[width=0.85\textwidth]{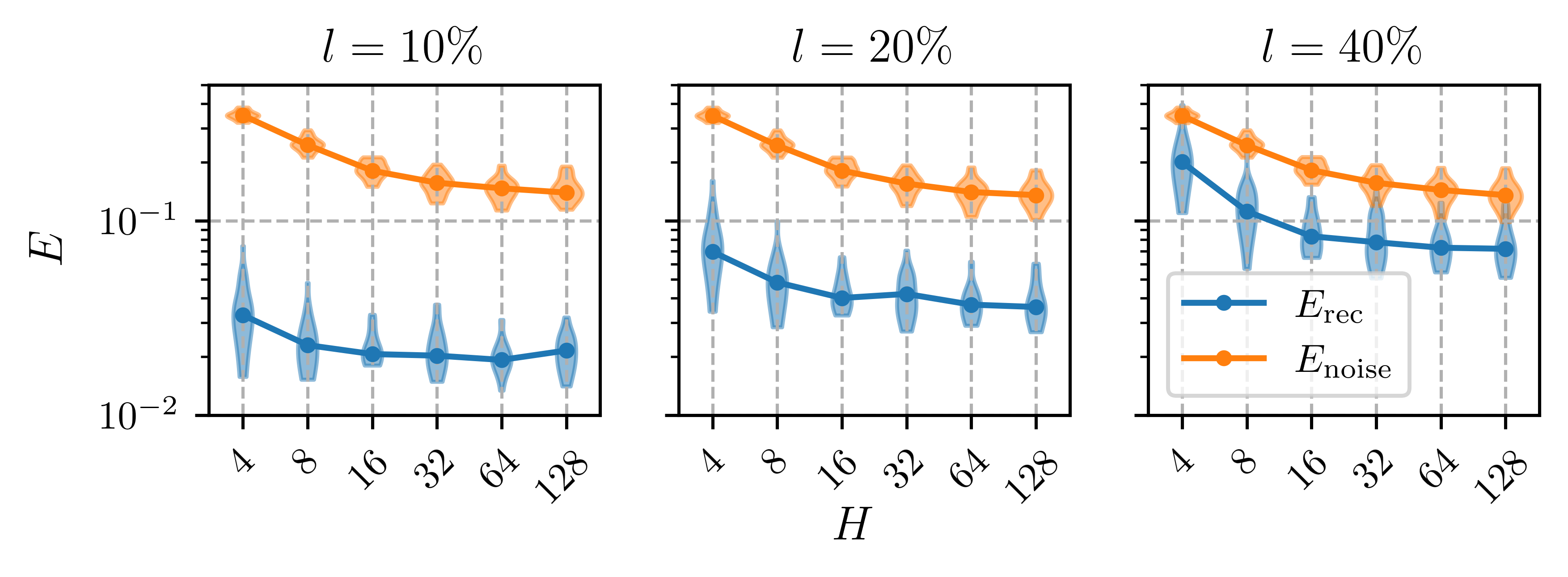}
		\caption{Influence of the prediction horizon $H$ on the noise \eqref{eq:e_noise} and coefficient \eqref{eq:e_rec} reconstruction errors; the markers in each violin plot indicate the median error over $20$ trials (varying noise samples).}
		\label{fig:test_horizon}
	\end{center}
\end{figure}

Figure \ref{fig:test_horizon} shows the influence of the prediction horizon on the identified noise and the reconstruction error.
Increasing $H$ improves the accuracy.
Beyond $H=32$, the accuracy remains relatively stable for all noise levels.
The relative improvement is more pronounced when the SNR is small.
Notably, more than $80\%$ of the randomly generated white noise is identified correctly if $H \ge 16$.

\begin{figure}[htbp]
	\begin{center}
		\includegraphics[width=0.85\textwidth]{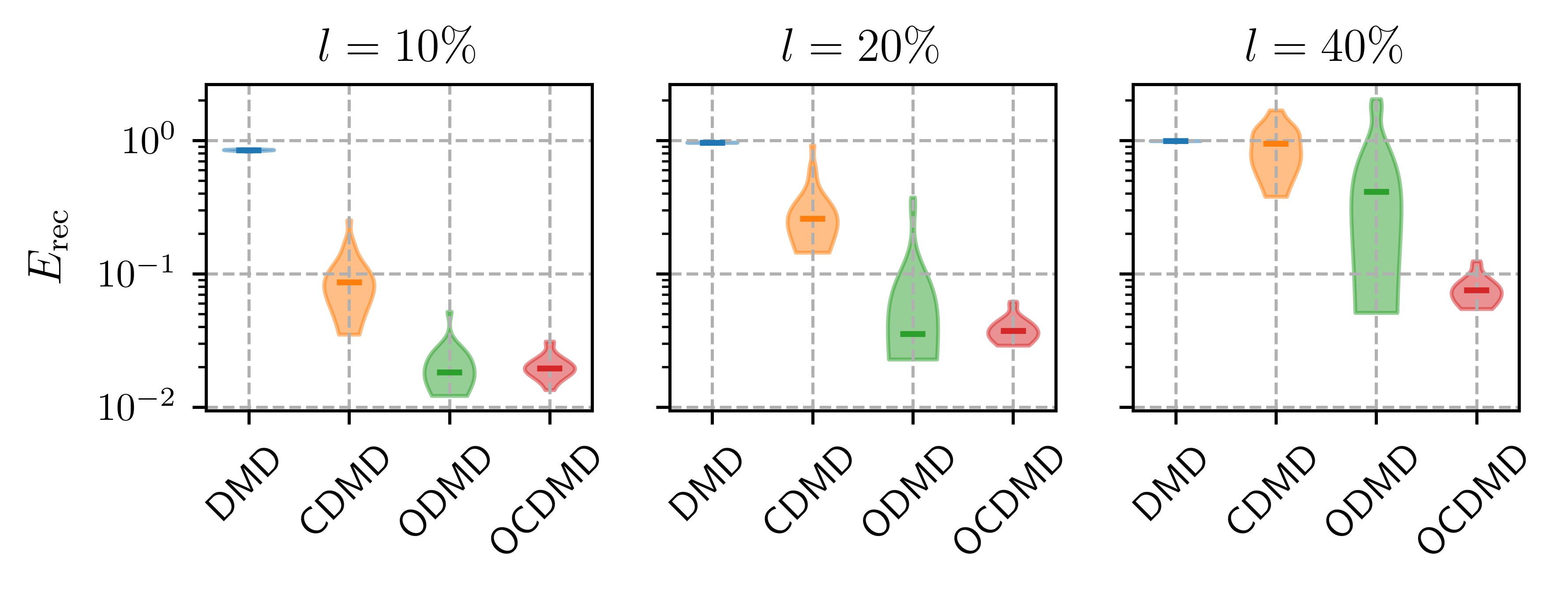}
		\caption{Comparison of the reconstruction error \eqref{eq:e_rec} achieved by different DMD variants over $20$ trials; the horizontal bar in each violin plot indicates the median reconstruction error; $H=64$ for the OCDMD.}
		\label{fig:test_models}
	\end{center}
\end{figure}

Figure \ref{fig:test_models} compares the DMD variants discussed in section \ref{sec:theory}.
We use the ODMD implementation available in PyDMD \cite{demo2018}.
All other variants are available in flowTorch \cite{weiner2021}.
The enhanced DMD variant introduced in this article is denoted OCDMD.
Based on the results presented in figure \ref{fig:test_horizon}, we set $H=64$.

The vanilla DMD's reconstruction error is close to $100\%$, since the noise corruption leads to quickly decaying dynamics.
The forward-backward consistency (CDMD) improves the results significantly for $l\leq 20\%$.
The reconstruction error is again close to $100\%$ for the highest noise level.
The median error of ODMD and OCDMD is similar for the high and middle SNR.
At the lowest SNR, the ODMD's accuracy breaks down, and the spread of the results grows considerably.
The OCDMD's prediction error increases with the noise level as well, but at a much lower rate. Even at the lowest SNR, the reconstruction error remains below $10\%$.
Moreover, the spread in the results is fairly small compared to CDMD and ODMD and remains nearly constant across the investigated noise levels.
A drawback of our approach is the increased computational cost.
However, considering the present data, the optimization finishes within less than a minute, and our current focus is on improving the robustness rather than the computational efficiency.

\section{Conclusion}
\label{sec:conclusion}

We introduced a flexible and robust approach to identifying linear dynamics and noise in flow data.
The approach combines several ideas from existing DMD variants and relies on automatic differentiation and gradient descent to solve the resulting optimization problem.
The results presented for the flow past a cylinder are promising in terms of accuracy and robustness.
Additional tests on more challenging datasets, e.g., flows with resolved turbulence or experimental data, are required to assess the proposed method more comprehensively, and we aim to present such tests in future research.
Algorithmically, the approach can be extended in various directions, e.g., to consider control inputs, parametric variations, or probabilistic modeling \cite{demo2018}.

\section*{Acknowledgment}
The authors gratefully acknowledge the Deutsche Forschungsgemeinschaft DFG (German Research Foundation) for funding this work in the framework of the research unit FOR 2895 under the grant WE 6948/1-1.
%
%
%
\printbibliography[title=References, heading=bibliography]

\end{document}